\documentclass[aps,pre,twocolumn,superscriptaddress,showpacs]{revtex4-2}
\usepackage{graphicx}
\usepackage{array}
\usepackage{amsfonts}
\usepackage{amssymb,amsmath,multirow,rotate}
\usepackage{mathrsfs}
\usepackage{booktabs}
\usepackage{threeparttable}
\usepackage{multirow}
\usepackage{epsfig}
\usepackage{threeparttable}
\usepackage{chngpage}
\usepackage{latexsym}
\usepackage{dcolumn}
\usepackage{graphics,graphicx,bm,fleqn,epic,eepic,float}
\usepackage{verbatim}
\usepackage{color}
\usepackage{xr}
\usepackage{float}
\usepackage{listings} % For including code in appendix
\usepackage{placeins} % For putting plots in the desired location
\usepackage{tikz}
\usepackage{amsmath, nccmath}
\usepackage{lipsum}  
\usepackage{soul}

\definecolor{red}{rgb}{1,0,0}

\definecolor{blue}{rgb}{0,0,1}

\begin{document}

\title{High-order exceptional points and stochastic resonance in pseudo-Hermitian systems}

\date{\today}

\author{Shirin Panahi}
\affiliation{School of Electrical, Computer, and Energy Engineering, Arizona State University, Tempe, AZ 85287, USA}

\author{Li-Li Ye}
\affiliation{School of Electrical, Computer, and Energy Engineering, Arizona State University, Tempe, AZ 85287, USA}

\author{Ying-Cheng Lai} \email{Ying-Cheng.Lai@asu.edu}
\affiliation{School of Electrical, Computer, and Energy Engineering, Arizona State University, Tempe, AZ 85287, USA}
\affiliation{Department of Physics, Arizona State University, Tempe, Arizona 85287, USA}

\begin{abstract}

Exceptional points, a remarkable phenomenon in physical systems, have been exploited for sensing applications. It has been demonstrated recently that it can also utilize as sensory threshold in which the interplay between exceptional-point dynamics and noise can lead to enhanced performance. Most existing works focused on second-order exceptional points. We investigate the stochastic dynamics associated with high-order exceptional points with a particular eye towards optimizing sensing performance by developing a theoretical framework based on pseudo-Hermiticity. Our analysis reveals three distinct types of frequency responses to external perturbations. A broad type of stochastic resonance is uncovered where, as the noise amplitude increases, the signal-to-noise ratio reaches a global maximum rapidly but with a slow decaying process afterwards, indicating achievable high performance in a wide range of the noise level. These results suggest that stochastic high-order exceptional-point dynamics can be exploited for applications in signal processing and sensor technologies.    

\end{abstract}

\maketitle

\section{Introduction} \label{sec:intro}

Non-Hermitian systems arise commonly in physical systems, quantum or classical. For 
example, when a quantum system is placed in an environment or a ``bath'' with mutual
interactions with the surroundings, the energy of the system is no longer 
conserved and it becomes non-Hermitian~\cite{cao2015dielectric,HXGL:2018,el2018non,kawabata2019symmetry,lee2020many,franca2022non}.
From a mathematical point of view, an imaginary part of the eigenenergy emerges,
making it complex~\cite{YHLG:2011,HYL:2011,YHLP:2012,NHLP:2012,YHLG:2012,WYLG:2013,NHYL:2013,YWHL:2014}. 
One approach to achieving real eigenenergies in a non-Hermitian system is by 
introducing pseudo-Hermiticity. Historically, the concept of pseudo-Hermiticity was 
first studied by Dirac and Pauli~\cite{pauli1943dirac} and was further developed by 
Mostafazadeh~\cite{mostafazadeh2002pseudo_1,mostafazadeh2002pseudo_2,mostafazadeh2002pseudo_3}, 
who exploited pseudo-Hermiticity to define general conditions for the real spectrum:
$UHU^{-1}=H^{\dagger}$, where $U$ is a linear Hermitian operator. For non-Hermitian 
quantum systems with energy dissipation, the fundamental governing equation is the 
master equation in the Lindblad 
formalism~\cite{lindblad1976generators,gorini1976completely}. Experimentally, the 
originally infinite sharp energy levels in the closed system become broadened, 
leading to various resonances with a finite width~\cite{gamow1928quantentheorie}, 
which are characteristic features of non-Hermitian quantum systems. Due to the 
equivalence between the Schr\"{o}dinger equation and the classical wave 
equations~\cite{kawano2004introduction,guidry2019three}, non-Hermiticity of 
different physical origins can also arise in a variety of classical systems, such 
as classical electric circuits with Joule 
heating~\cite{joglekar2009elusive,xiao2019enhanced,yin2023high,li2023stochastic} 
and photonics devices with gain and loss of photons described by the Maxwell 
equations~\cite{kononchuk2020orientation,xiong2021higher}.

A remarkable phenomenon in non-Hermitian physical systems is the occurrence of 
exceptional points (EPs)~\cite{heiss2012physics,kato2013perturbation,miri2019exceptional}.
An EP is the point in the parameter space at which $n$ ($n \ge 2$) eigenvalues of the 
non-Hermitian Hamiltonian matrix and their eigenstates coalesce and emerge as a 
branch-point singularity. A low-order or second-order EP is referred to as $n = 2$, 
while $n \ge 3$ defines a high-order EP. An EP differs from a diabolic point in 
conventional spectral analysis at which some eigenvalues are degenerate, but not 
their associated eigenstates. In sensor applications based on photonic, acoustic, or 
electronic systems exhibiting resonances~\cite{hodaei2017enhanced,zhao2018exceptional,xiao2019enhanced,kononchuk2022exceptional}, 
EPs can be exploited to amplify the sensitivity~\cite{wiersig2014enhancing} of the 
sensor, a feat that cannot be achieved through a diabolic point. In particular, for 
a diabolic point, when a perturbation or a weak signal of the strength 
$\varepsilon \ll 1$ is applied to the system, the splitting of the energy or frequency 
measured in the transmission or reflection spectrum is proportional to $\varepsilon$. 
However, for an EP, the energy splitting can be $\varepsilon^{1/(n-1)}$. The higher 
the order of an EP, the more sensitive the sensor response. The enhancement of the 
system response about an EP provides a physical mechanism to optimize the sensitivity 
of the sensors based on detecting the frequency splitting, such as microcavity 
sensors~\cite{wiersig2014enhancing,vollmer2012review}, optical 
gyroscopes~\cite{sunada2007design}, weak magnetic field sensors~\cite{rondin2014magnetometry,barry2020sensitivity}, and nanomechanical mass sensors~\cite{djorwe2019exceptional,chaste2012nanomechanical,gil2010nanomechanical}. 

In the study of EPs, the symmetries of the system play an important 
role~\cite{sep-symmetry-breaking}. In non-Hermitian physical systems, the 
parity-time (PT) symmetry is fundamental. For example, in a hard-core Bose gas modeled 
by the Toda lattice~\cite{hollowood1992solitons} with the Fermi's 
pseudopotential~\cite{wu1959ground}, the PT symmetry can produce the entire real 
spectrum in the non-Hermitian matrices~\cite{BPhysRevLett}. A spontaneous 
PT symmetry breaking~\cite{ashida2020non,yuan2020steady} occurs when some eigenstates 
of the Hamiltonian are not the eigenstates of the PT operator, making some pairs of 
eigenvalues complex conjugate to each other and generating the PT-broken phase. In 
this case, a second-order EP emerges between the PT-exact and PT-broken phases. The 
low-order EP with PT symmetry can be realized in a variety of physical and engineering 
systems, e.,g., as gain-loss structures in optics~\cite{bergman2021observation,xiao2017observation,peng2014parity,ruter2010observation}, electronics~\cite{yang2022observation,assawaworrarit2017robust,chen2018generalized}, microwaves~\cite{liu2018observation}, mechanics and acoustics~\cite{shi2016accessing,fleury2015invisible}, superconducting 
circuits~\cite{quijandria2018pt,zhang2019dispersive,naghiloo2019quantum}, 
as well as spin~\cite{wu2019observation} and atomic systems~\cite{zhang2016observation}. 
These systems provide rich experimental settings for investigating various phenomena 
associated with low-order EPs, such as unidirectional 
invisibility~\cite{lin2011unidirectional,chang2014parity}, sensitivity 
measurement~\cite{chen2017exceptional,hokmabadi2019non}, and Berry phase induced from 
encircling EPs~\cite{dembowski2001experimental,lee2009observation,gao2015observation,el2018non}. 

While high-order EPs exhibit merits over low-order EPs such as 
sensitivity~\cite{hodaei2017enhanced,zeng2021ultra,wang2021enhanced}, spontaneous 
emission enhancement~\cite{lin2016enhanced}, and certain topological 
characteristics~\cite{ding2016emergence,delplace2021symmetry,mandal2021symmetry},
experimental implementation of high-order EPs with the PT symmetry is 
challenging~\cite{yin2023high,SK2022}. For example, to generate a third-order EP, 
in between gain and loss resonators, a neutral resonator is 
required~\cite{wiersig2020review,yin2023high}. To realize such a device, completely 
lossless and gainless features are needed, with equal coupling between the adjacent 
resonators. Accordingly, pseudo-Hermitian structures without the PT symmetry was 
proposed~\cite{xiong2021higher,yin2023high} to generate high-order EPs in a cavity 
optomechanical system, which only requires gain and loss elements. Such a design 
somewhat relaxes the constraint and provides more freedom in the design of high-order 
EP based sensors. With the state-of-the-art experimental technologies, high-order 
EPs in pseudo-Hermitian systems have begun to be realized in cavity 
magnonics~\cite{ZPhysRevB}, cavity optomechanical 
systems~\cite{xiong2022higher,xiong2021higher,yin2023high}, photonic Lieb 
lattices~\cite{xia2021higher}, and the Su-Schrieffer-Heeger 
chain~\cite{barnett2023locality}.

The interplay between noise and EP dynamics has attracted recent attention. The 
phenomenon of stochastic EP was uncovered recently in a sensor system with the PT 
symmetry and a second-order EP~\cite{li2023stochastic}. In particular, with a weak 
periodic signal as the input, the sensory threshold fluctuates with random PT-phase 
transitions, suggesting that noise can serve to enhance the sensor performance. 
This work~\cite{li2023stochastic} thus built up a bridge between EP dynamics and the 
paradigmatic and extensively studied phenomenon of stochastic 
resonance~\cite{benzi1981mechanism,gammaitoni1998stochastic}, where noise can be 
used to optimize the system's response to weak signals in terms of measures such 
as the signal-to-noise ratio (SNR). So far, this connection has been explored for 
second-order EP systems with the PT symmetry.

In this paper, we investigate stochastic high-order EP dynamics in pseudo-Hermitian 
systems. To make feasible experimental implementation, we generalize 
the necessary conditions for stochastic low-order EP~\cite{li2023stochastic} in terms 
of the complex relation between the system's eigenfrequencies and external disturbance 
by introducing a more general physical framework for high-order EPs. We identify three 
distinct types of frequency responses of the system to perturbations, suggesting that 
stochastic high-order EP can be exploited for designing highly sensitive and robust 
sensors. Our main finding is the emergence of a remarkable skewed stochastic resonance: 
as the noise amplitude increases, the SNR increases and reaches the maximum rapidly, 
followed by a significantly slow decay. The implication is that a wide range of the 
noise level can be used to achieve the optimal or near optimal SNR. While this is akin 
to the phenomenon of stochastic resonance without tuning~\cite{CCI:1995,CKFW:2001}, we 
note that the latter typically occurs in spatially extended dynamical systems that 
presents a challenge for experimental implementation as sensors. To our knowledge, our 
stochastic high-order EP system represents a class of non-spatially-extended systems 
in which an extensive stochastic resonance can arise.

A relevant question is how the SNR depends on the signal amplitude. To address
this question, we note that a purpose of generating an EP is for its use as
a sensory threshold in scenarios that involve weak periodic signals. About an
EP, the system can be highly sensitive to small perturbations, and this
sensitivity plays a crucial role in the SNR. In particular, the sensor can be
so designed that the weak periodic input signal is ``at'' or ``below'' the
sensory threshold. This means that, in the absence of noise, the signal would
not be strong enough to produce an output response. However, when noise is
introduced into the system, it can induce random transitions, thereby enhancing
the sensor's performance. We thus have the characteristics of the phenomenon of
a stochastic resonance, where noise can ``push'' the signal beyond the threshold, 
allowing an output event to be detected. The noise strength is critical, as it 
determines how frequently and by how much the signal exceeds the threshold, 
thus impacting the SNR. While the SNR depends on the noise strength, the 
relationship also hinges on the interplay between noise and the signal amplitude, 
with certain noise levels optimizing sensor performance and maximizing the SNR.
In particular, the dependency of the SNR on the signal amplitude is closely tied 
to the noise level, especially before or at an EP that acts as a sensory threshold. 
For signals with such an amplitude, the output amplitude increases with the 
rising noise level. However, this does not necessarily improve the output signal. 
If the noise level surpasses a certain threshold, the output becomes dominated by 
noise, thereby losing the meaningful signal information. Overall, the optimal 
conditions for maximizing the SNR are contingent upon maintaining a balanced 
noise level that is sufficient to enhance the signal without overwhelming it and 
causing degradation.

\section{High-order exceptional points} \label{sec:theory}

\subsection{Basics}

\begin{figure} [ht!]
\centering
\includegraphics[width=\linewidth]{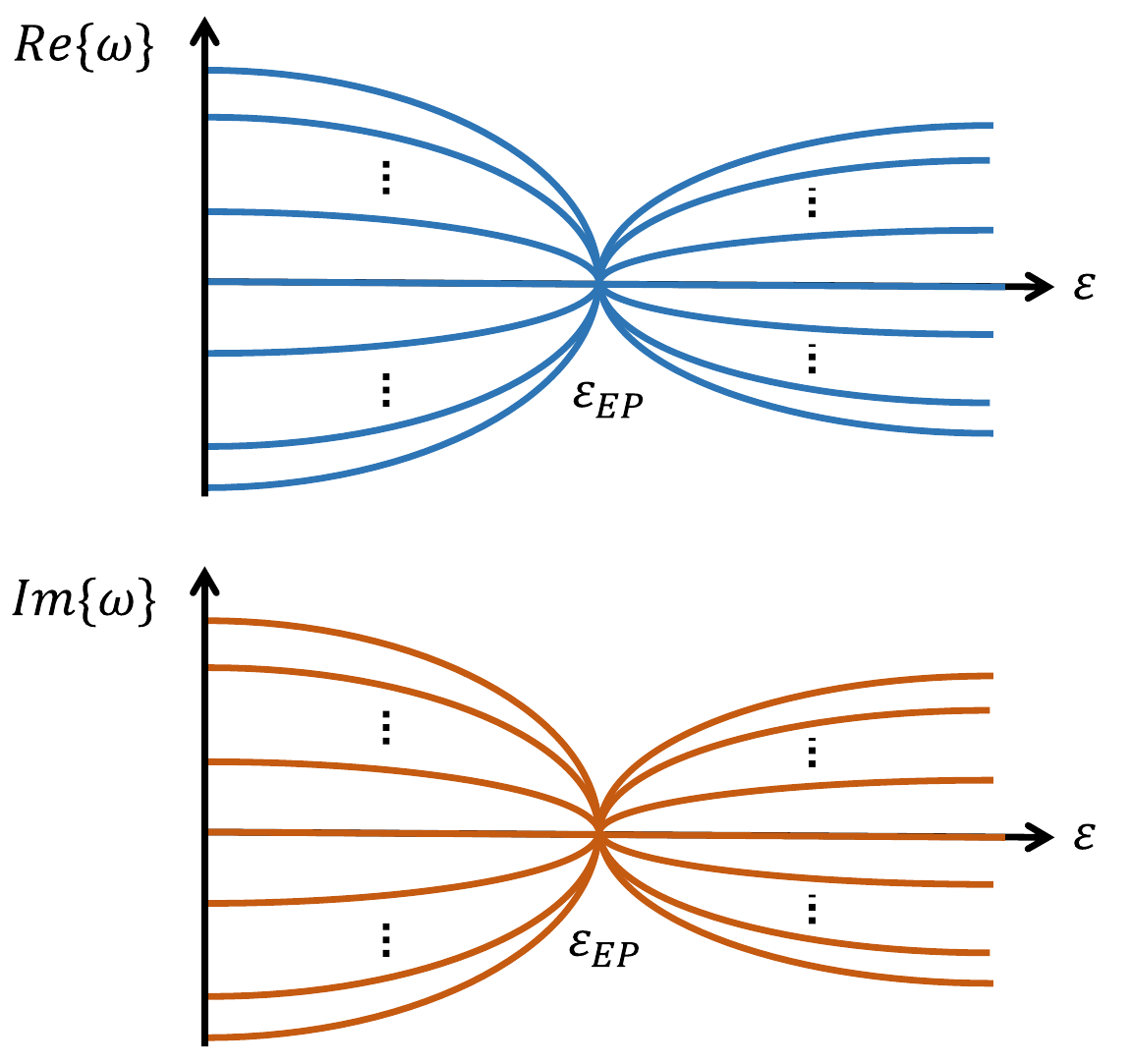}
\caption{Schematic illustration of the real (top) and imaginary (bottom) components 
of the eigenvalues as a function of the perturbation $\varepsilon$. The critical value of
the perturbation at which the eigenvalues and corresponding eigenvectors of the system 
coalesce is $\varepsilon_{EP}$.}
\label{fig:EP_schematic}
\end{figure}

The types of physical systems in which EPs arise are generally open and are 
described by a non-Hermitian Hamiltonian. To be concrete, we consider a system 
represented by an $n \times n$ effective Hamiltonian matrix $H_0$ and a parameter 
$\varepsilon$ characterizing some perturbation $\hat{H}$ to the system. The total 
Hamiltonian is $H = H_0 + \hat{H}(\varepsilon)$, whose eigenvalues determine the 
system's response to the perturbation. The resultant eigenvalue 
splittings represent measurable output quantities accessible through a spectral 
analysis. Figure~\ref{fig:EP_schematic} schematically illustrates the real and 
imaginary parts of the eigenvalues as a function of $\varepsilon$ for a system with 
EP. In particular, because of the non-Hermitian nature of the Hamiltonian, the 
eigenvalues $\omega$ are generally complex, where the real and imaginary parts 
characterize the frequency and the linewidth of the system response, respectively. 
For values of $\varepsilon$ in an open interval, the eigenvalues are typically 
distinct. An EP is the value of $\varepsilon$ at which both the real and imaginary 
parts of the eigenvalues coalesce, as shown in Fig.~\ref{fig:EP_schematic}.

Let ${\rm Re}\{\delta\omega\}$ and ${\rm Im}\{\delta\omega\}$ be the real and imaginary 
parts of the eigenvalue splittings, respectively. The $Q$ factor of the system, defined as
\begin{align} \nonumber
Q = {\rm Re}\{\delta\omega\}/(2{\rm Im}\{\delta\omega\}), 
\end{align}
is often used as an index in resonator circuits, which indicates the sharpness of the 
dip within the measured reflection-spectrum curve. In general, a larger real part and/or 
a smaller imaginary part of the eigenfrequency spitting can lead to a larger $Q$ factor, 
yielding higher spectral resolution. Experimentally, the eigenfrequency with the smallest
imaginary part corresponds to the sharpest detectable reflection spectral dip. 

\subsection{Scenarios for emergence of high-order exceptional points}

As the magnitude of the perturbation changes, there are three common scenarios through
which an EP can arise: branch, monotonic, and non-injective.

\begin{figure}[ht!]
\centering
\includegraphics[width=\linewidth]{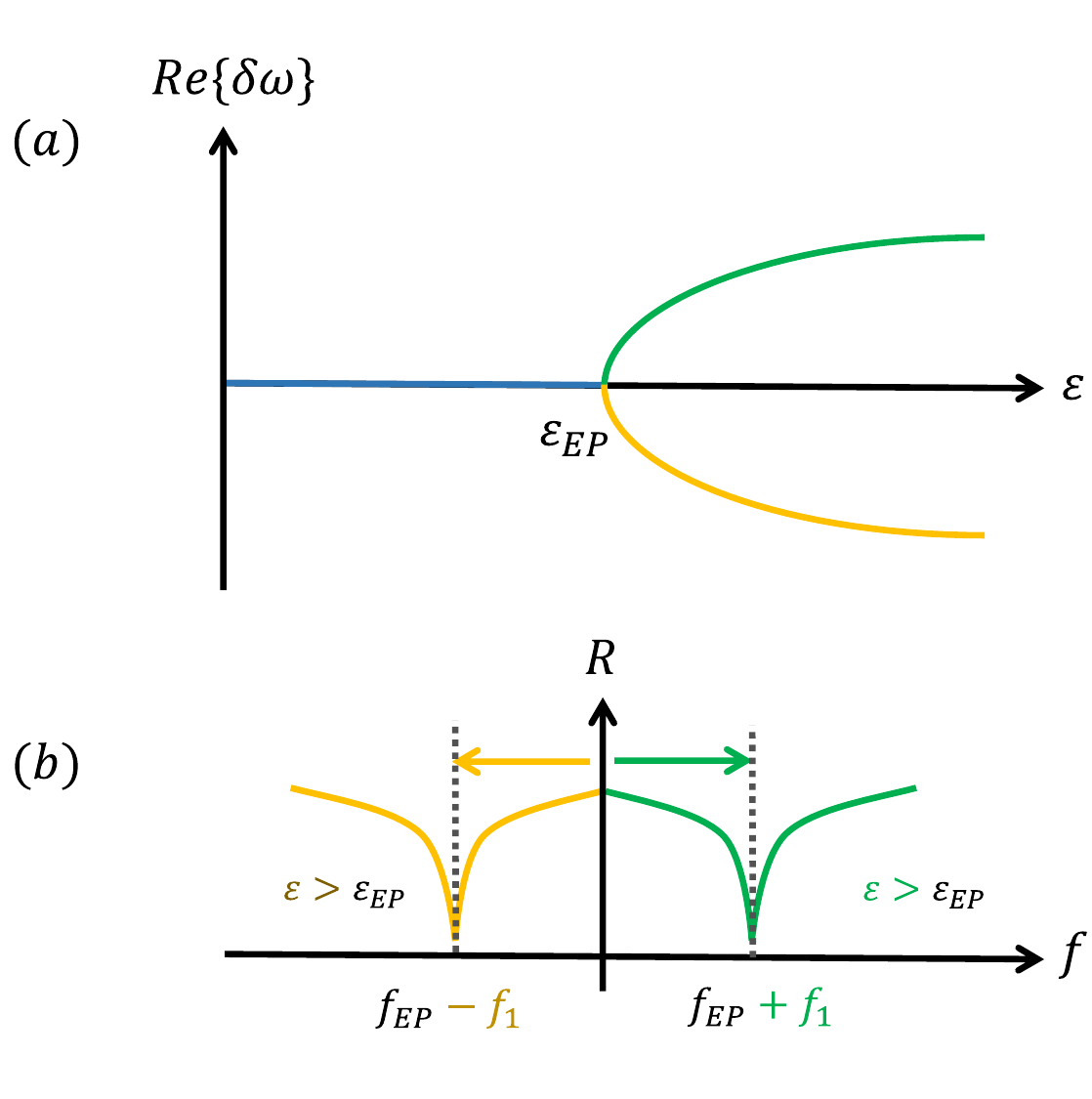}
\caption{Emergence of an EP through a branch structure. (a) Real part of frequency 
splitting ${\rm Re}\{\delta \omega \}$ versus the magnitude $\varepsilon$ of the 
perturbation, where an EP arises at $\varepsilon_{EP}$. (b) Consequence of the EP: for 
$\varepsilon \agt\varepsilon_{EP}$, a pair of dips in the reflection coefficient are 
created at $f_{EP} \pm f_1$, where $f_1$ is the absolute value of 
${\rm Re}\{\delta\omega\}$ at $\varepsilon \agt\varepsilon_{EP}$ in (a).}
\label{fig:branch}
\end{figure}

\paragraph{Branch scenario.}
This scenario arises when the system exhibits a branch-like response to the 
perturbation, as illustrated in Fig.~\ref{fig:branch}. At the critical bifurcation
point $\varepsilon_{EP}$, all the eigenvalues and their corresponding eigenvectors 
coalesce. For $\varepsilon < \varepsilon_{EP}$, the real part of the splitting of the 
eigenvalue with the smallest imaginary part, ${\rm Re}\{\delta\omega\}$, is constant. 
For $\varepsilon>\varepsilon_{EP}$, ${\rm Re}\{\delta\omega\}$ has two possible values 
with opposite signs, which are symmetric to each other with respect to the 
$\varepsilon$-axis, as shown in Fig.~\ref{fig:branch}(a).

To give a physical example, consider an open cavity that reflects and transmits 
an incoming wave and the quantities of interest are the reflection and transmission
coefficients as a function of the frequency. A small positive deviation from the 
critical point $\varepsilon_{EP}$ toward the right will simultaneously lead to a 
positive and negative shift in the frequency: $\pm {\rm Re}\{\delta\omega\}$, giving 
rise to two dips in the reflection coefficient at $f_{EP} \pm f_1$, respectively, as
shown in Fig.~\ref{fig:branch}(b). Accordingly, at each of the two dips, the 
transmission coefficient exhibits a peak. 

\begin{figure} [ht!]
\centering
\includegraphics[width=\linewidth]{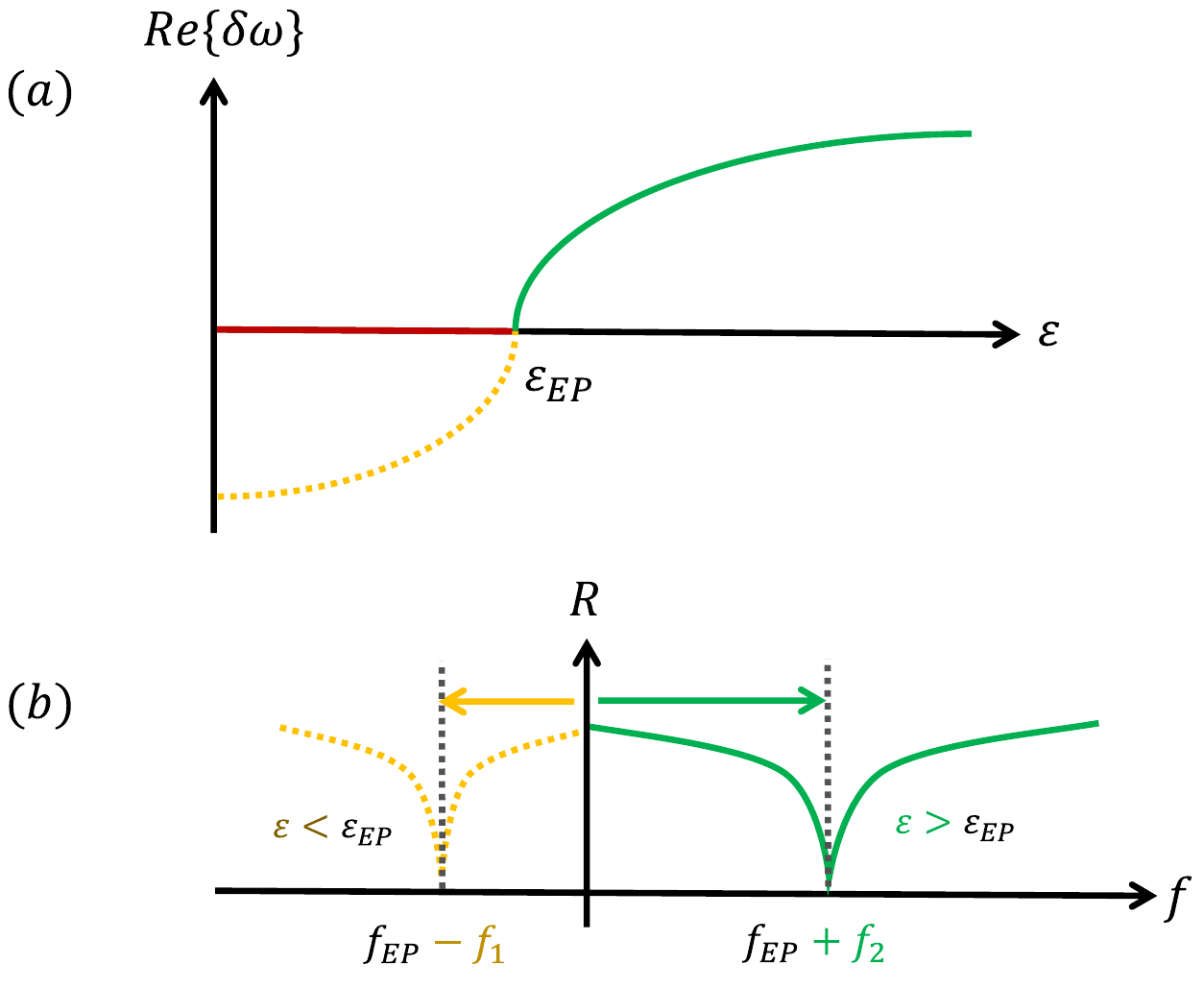}
\caption{Monotonic scenario. (a) Real frequency shift ${\rm Re}\{\delta \omega \}$ 
versus the perturbation about the critical value $\varepsilon_{EP}$. Depending on 
whether $\varepsilon$ is below or above $\varepsilon_{EP}$, the amount of the shift is
negative or positive, respectively. (b) The resulting dips in the reflection 
coefficient, one for $\varepsilon < \varepsilon_{EP}$ and another for 
$\varepsilon > \varepsilon_{EP}$. Frequency filtering can be employed to remove a dip.}
\label{fig:monotonic}
\end{figure}

\paragraph*{Monotonic scenario.}
In this scenario, as the perturbation is strengthened, the system's frequency response
changes monotonically, as illustrated in Fig.~\ref{fig:monotonic}, where there is a 
critical saddle point at $\varepsilon_{EP}$. For $\varepsilon$ deviating from 
$\varepsilon_{EP}$, the real part of the eigenvalue with the smallest imaginary part does
not split but shifts by a small amount with its sign depending on whether the 
perturbation is smaller or larger than $\varepsilon_{EP}$. In particular, for  
$\varepsilon \alt \varepsilon_{EP}$, the frequency shift ${\rm Re}\{\delta \omega \}$ is 
negative but it is positive for $\varepsilon \agt \varepsilon_{EP}$, as shown in 
Fig.~\ref{fig:monotonic}(a). In this case, a perturbation smaller or larger than 
$\varepsilon_{EP}$ result in two distinct values of ${\rm Re}\{\delta \omega\}$ with 
opposite signs, leading to two asymmetric dips in the spectrum of the reflection 
coefficient, one below and another above $f_{EP}$, as shown in 
Fig.~\ref{fig:monotonic}(b). Because of the monotonic behavior of 
${\rm Re}\{\delta \omega \}$ with respect to variations of the perturbation, it is 
possible to extract the system response from one side of the EP. For example, a filter 
can be employed to remove the frequency response of the system to perturbations smaller 
than $\varepsilon_{EP}$, as shown by the dash-dotted curve in Fig.~\ref{fig:monotonic}(a),
leaving the frequency response unchanged in spite of the perturbation (represented by 
the red solid line). 

\begin{figure} [ht!]
\centering
\includegraphics[width=\linewidth]{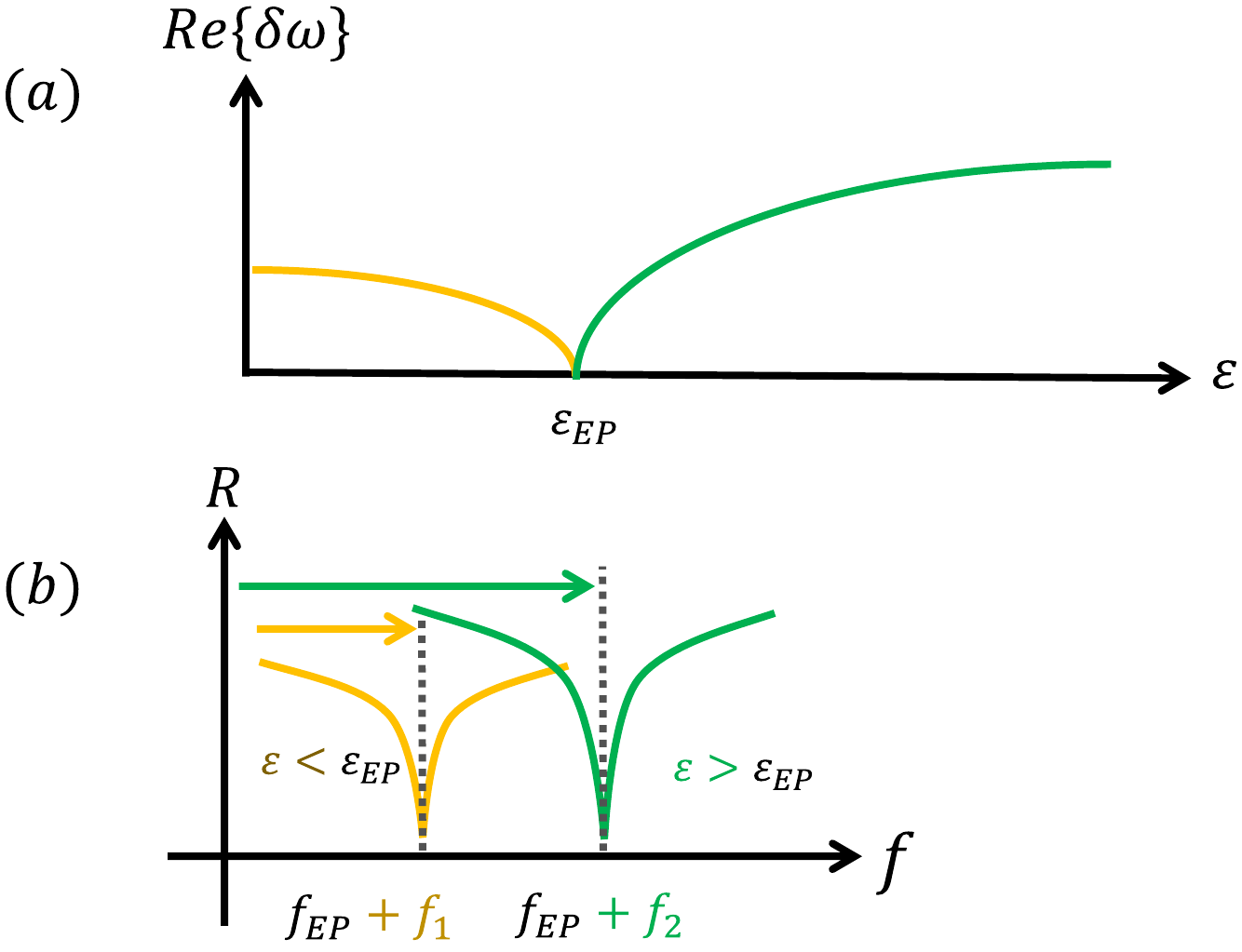}
\caption{A non-injective structure. (a) Real frequency splitting 
${\rm Re}\{\delta\omega\}$ for different perturbations. (b) A schematic illustration 
of the response of a system in terms of the reflection coefficient under a 
perturbation. Any perturbation leads to a frequency shift to the right. An 
example for $\varepsilon < \varepsilon_{EP}$ ($\varepsilon > \varepsilon_{EP}$) is 
shown in yellow (green).}
\label{fig:noninjective}
\end{figure}

\paragraph*{Non-injective Structure.}
Figure~\ref{fig:noninjective} shows the non-injective structure of the system 
response to the applied perturbation with respect to an EP or the bifurcation point 
at $\varepsilon_{EP}$. The system lacks the injective properties, i.e., there is no 
one-to-one mapping between the distinct elements of its output domain and those of 
the input domain. In particular, Fig.~\ref{fig:noninjective}(a) illustrates that the 
system's response is nearly symmetric around the EP. Figure~\ref{fig:noninjective}(b) 
schematically illustrates two examples of the response of the system in terms of the 
reflection coefficient, which indicate a rightward shift in both cases. As a result, 
distinguishing between the response corresponding to $\varepsilon<\varepsilon_{EP}$ 
(yellow) and $\varepsilon>\varepsilon_{EP}$ (green) becomes infeasible.

In general, in a non-injective structure, the response of the system is approximately
symmetric with respect to the EP, so both larger and smaller perturbations can cause
a similar shift in the frequency spectrum. As a result, it becomes difficult to
distinguish whether a particular frequency shift is caused by a perturbation below
or beyond the EP, complicating the system's use as a reliable threshold sensor.
Essentially, the system is unable to provide a directional response, i.e., to
respond only when the perturbation exceeds a certain threshold value, because
perturbations on either side of the EP would produce a similar spectral shift.
A concrete example (a PT-symmetric electrical circuit based on a sixth-order
EP) is provided in Sec.~\ref{sec:example1}.

It is worth emphasizing that, for a system designed to function as a threshold sensor,
it is critical that the structure is injective, meaning that the response
is monotonic and distinct in one direction only, either for perturbations above or
below the threshold. An EP-based sensor intended to function as a sensory threshold
should have a structure where the response is confined to one branch, ensuring that
only perturbations above (or below) the EP elicit a detectable output. In a
non-injective structure, this condition is not met, and the sensor becomes unsuitable
for threshold applications.

\subsection{Stochastic resonance}

Stochastic resonance is a fundamental phenomenon in nonlinear and statistical 
physics~\cite{gammaitoni1998stochastic,park2007noise,lai2009stochastic,ying2018enhancing},
where an optimal level of noise can have a beneficial role in enhancing the system’s 
response to weak inputs. Recently, stochastic processes in EP-based structures 
were investigated in~\cite{li2023stochastic}, opening the door to innovative sensor 
designs capable of exploiting the inherent noise of the system to achieve improved 
performance with implications across diverse application domains. 
 
In the theory of stochastic processes, the EPs in certain structures exhibit a unique 
capacity: they can operate as dynamic sensory thresholds, generating random frequency 
shifts when subject to a time-varying input. In these systems, EPs play a role in 
separating two distinct phases of the system response, one which is insensitive to 
the input (${\rm Re}\{\delta \omega \}=0$) and another that is sensitive 
${\rm Re}\{\delta \omega \}>0$. Further, the system is able to exploit the fluctuations 
induced by noise to its advantage, thereby enhancing its sensitivity to weak input 
variations, a stochastic-resonance like phenomenon. To quantify a stochastic resonance 
in a system with an EP, we analyze whether the additional noise paradoxically increases 
the sensor's SNR. Our aim is to demonstrate that a stochastic resonance around the EPs 
(stochastic EPs) can be exploited for developing sensors capable of functioning 
effectively under environmental fluctuations. 

It is worth noting that the primary difference between a second-order EP and a
higher order EP lies in their sensitivity to perturbations. For the latter, such as 
a third-order EP, the system's sensitivity can increase significantly compared to a
second-order EP. This sensitivity can be quantified by the relationship 
$\epsilon^{1/(n-1)}$, where $n$ denotes the order of the EP. Consequently, a
system operating at a third-order EP is more sensitive to noise than one at a 
second-order EP, meaning that even smaller perturbations can induce a larger response. 
Practically, this enhanced sensitivity can lead to a steeper initial rise in the SNR 
as the noise amplitude ($\sigma$) increases, because the system is more responsive to 
stochastic fluctuations. However, due to this heightened sensitivity, a higher-order 
EP system reaches the critical noise level sooner, beyond which the SNR begins to 
decrease. This contrasts with a second-order EP, where the system is less sensitive 
to perturbation, causing the critical noise level to occur at a higher amplitude. 
Overall, while both second-order and higher-order EPs show an increase in the SNR with 
noise up to a critical value followed by a decline, a system with a higher-order EP 
may reach its peak SNR earlier and is more sensitive to small noise perturbations.

To offer a more comprehensive understanding of the three EP structures, we 
study each within the context of electrical sensors through case studies. 

\section{Stochastic resonance in systems with high-order exceptional points} \label{sec:example1}

\begin{figure}[h]
\centering
\includegraphics[width=\linewidth]{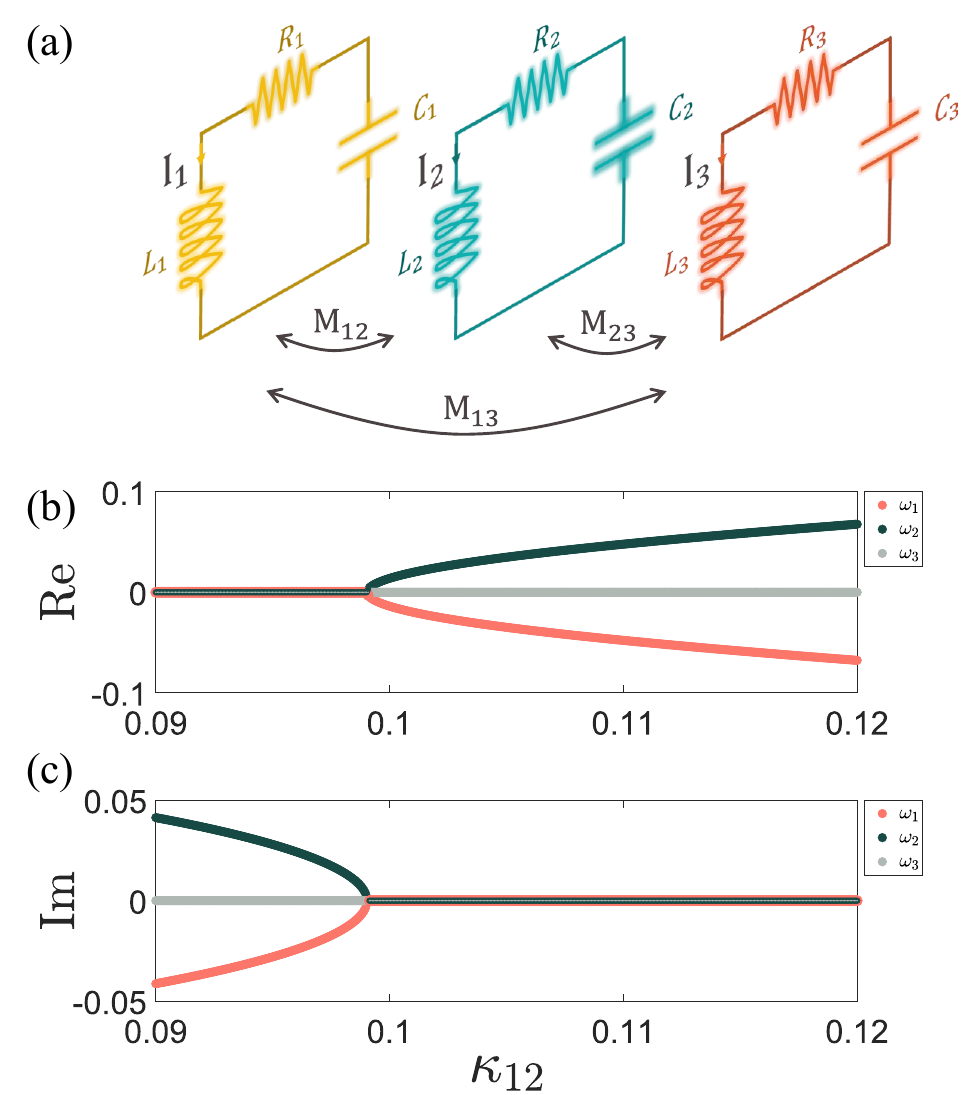}
\caption{System of three inductively coupled resonators as a wireless sensor. (a) A 
schematic illustration of the circuit system. (b,c) Real and imaginary parts of the 
eigenfrequency of the system~\eqref{Eq: schrodinger} as a function of mutual coupling 
parameter $\kappa_{12}$, respectively. Other parameter values are $\alpha=50$, 
$\gamma = 0.1$, $\kappa_{13} = 0$, and $\kappa_{23} = (1+\alpha)^{3/2}\kappa_{12}$.
The response of the three coupled RLC resonators exhibits a branch structure: for 
$\kappa_{12} < \kappa_{EP}$, the perturbed system has one real eigenfrequency 
($\omega_3 = 0$) and two complex conjugate eigenfrequencies, whereas for 
$\kappa_{12} > \kappa_{EP}$, it has three distinct real eigenfrequencies.}
\label{fig:coupled_system}
\end{figure}

Our prototypical system consists of three inductively coupled resonators as a wireless 
sensor~\cite{yin2023high}, which are described by a pseudo-Hermitian effective 
matrix, as shown in Fig.~\ref{fig:coupled_system}(a). The sensor can be designed to 
exhibit a high-order EP in both branch and monotonic structures, thereby capable of 
operating as a dynamic sensory threshold. The system equations can be derived by a 
standard circuit analysis. In particular, the three inductors are coupled through 
electromagnetic induction in which the voltages $V_{L_n}$ and currents $I_n$ flowing 
through the inductors are related to each other as
\begin{equation}
    \begin{bmatrix}
        V_{L_1}\\
        V_{L_2}\\
        V_{L_3}
    \end{bmatrix} = i\omega
    \begin{bmatrix}
        L_1    &   M_{12} &  M_{13}\\
        M_{12} &    L_2   &  M_{23}\\
        M_{13} &   M_{23} &    L_3
    \end{bmatrix}
    \begin{bmatrix}
        I_1\\
        I_2\\
        I_3
    \end{bmatrix},
\end{equation}
where $L_n$ are the inductances and $M_{nk}$ ($n,k=1,2,3$) is the mutual inductance. 
Applying the Kirchhoff voltage law, we have
\begin{subequations}\label{Eq:kvl}
\begin{equation}
\begin{medsize}
            i\omega I_1 +  i\omega \frac{M_{12}}{L_1} I_2 +  i\omega \frac{M_{13}}{L_1} I_3 + \frac{R_1}{L_1} I_1 + \frac{1}{ic_1 L_1 \omega} I_1 = 0,
\end{medsize}
\end{equation}
\begin{equation}
\begin{medsize}
        i\omega \frac{M_{12}}{L_2} I_1 + i\omega I_2 + i\omega \frac{M_{23}}{L_2} I_3 + \frac{R_2}{L_2} I_2 + \frac{1}{ic_2 L_2 \omega} I_2=0,
\end{medsize}
\end{equation}
\begin{equation}
\begin{medsize}
   i\omega \frac{M_{13}}{L_3} I_1 + i\omega \frac{M_{23}}{L_3} I_2 + i\omega I_3 + \frac{R_3}{L_3} I_3 + \frac{1}{ic_3 L_3 \omega} I_3=0.
\end{medsize}
\end{equation}
\end{subequations}
These equations can be recast into a Schr\"{o}dinger-type of equation:
\begin{subequations} \label{Eq: schrodinger}
\begin{equation}
       i\dot{I} = H I,
       \end{equation}
       \begin{equation}
         \begin{medsize}
             H =  \begin{bmatrix}
        -i\frac{R_1}{L_1} - \frac{1}{C_1 L_1 \omega} & \frac{M_{12}}{L_1} \omega & \frac{M_{13}}{L_1} \omega\\[0.3cm]
        \frac{M_{12}}{L_2} \omega &  -i\frac{R_2}{L_2} - \frac{1}{C_2 L_2 \omega} & \frac{M_{23}}{L_2}\\[0.3cm]
        \frac{M_{13}}{L_3} \omega & \frac{M_{23}}{L_3} & -i\frac{R_3}{L_3} - \frac{1}{C_3 L_3 \omega}
    \end{bmatrix},
         \end{medsize}  
       \end{equation}
\end{subequations}
where $I=(I_1, I_2, I_3)^T$ are the system variables and $H$ is the effective 
Hamiltonian operator. The simplified approximation of Eq.~\eqref{Eq: schrodinger} 
can be obtained by assuming $\omega \approx \omega_0$ and defining the resonant 
frequencies as $\omega_0 = 1/\sqrt{C_nL_n}$ with $L_n = L$ and $C_n = C$. The 
gain/loss parameter is $\gamma_n=R_n\sqrt{C/L}$ and the inductive coupling 
coefficient is $\kappa_{nk}=M_{nk}/L$. The effective Hamiltonian is then 
reduced to $H = \omega_0(\tilde{H} - \mathbb{I})$:
\begin{equation}
\tilde{H} =
    \begin{bmatrix}
        -i\gamma_1 & \kappa_{12} & \kappa_{13}\\
        \kappa_{12} & -i\gamma_2 & \kappa_{23}\\
        \kappa_{13} & \kappa_{23} & -i\gamma_3
    \end{bmatrix},
\end{equation}
where $\mathbb{I}$ is a $3 \times 3$ identity matrix. Using the substitution 
$\tilde{\omega}=1+\omega/\omega_0$, we can calculate the eigenfrequencies through 
the associated characteristic equation 
$\det{\left(\tilde{H}-\tilde{\omega} \mathbb{I}\right)}=0$:
\begin{align}
\begin{split}
           &\tilde{\omega}^3 +i\left(\gamma_1+\gamma_2+\gamma_3\right)\tilde{\omega}^2\\
     &-\left(\kappa_{12}^2+\kappa_{13}^2+\kappa_{23}^2+\gamma_1\gamma_2+\gamma_1\gamma_3+\gamma_2\gamma_3\right)\tilde{\omega}\\ 
    & -\left(2\kappa_{12}\kappa_{13}\kappa_{23}+i\left(\gamma_1\gamma_2\gamma_3+\gamma_1\kappa_{23}^2+\gamma_2\kappa_{13}^2+\gamma_3\kappa_{12}^2\right)\right)=0.
    \end{split}
\end{align}
The real and imaginary parts of the characteristic equation are, respectively,
\begin{align}\label{Eq: Reldet}
	\tilde{\omega}^3 &-\left(\kappa_{12}^2+\kappa_{13}^2+\kappa_{23}^2+\gamma_1\gamma_2+\gamma_1\gamma_3+\gamma_2\gamma_3\right)\tilde{\omega}\\ \nonumber
        &-\left(2\kappa_{12}\kappa_{13}\kappa_{23}\right)=0,
\end{align}
and
\begin{align}\label{Eq: Imdet}
            &\left(\gamma_1+\gamma_2+\gamma_3\right)\tilde{\omega}^2\\ \nonumber
            &-\left(\gamma_1\gamma_2\gamma_3+\gamma_1\kappa_{23}^2+\gamma_2\kappa_{13}^2+\gamma_3\kappa_{12}^2\right)=0.
\end{align}
From the characteristic equation, we can then determine the conditions for the system 
to be pseudo-Hermitian and the conditions under which a third-order EP can emerge. 

\paragraph*{Pseudo-Hermiticity.}
A system whose Hamiltonian can be related to its adjoint through a similarity 
transformation is a pseudo-Hermitian system satisfying
\begin{align} \nonumber
\det{\left(\tilde{H}-\tilde{\omega} I\right)}=\det{\left(\tilde{H}^\dag-\tilde{\omega} I\right)}. 
\end{align}
For a symmetric Hamiltonian, this condition can be simplified to 
\begin{align} \nonumber
\det{\left(\tilde{H}-\tilde{\omega} I\right)}=\det{\left(\tilde{H}^*-\tilde{\omega} I\right)}
\end{align}
or
\begin{align} \nonumber
{\rm Im}\{\det{\left(\tilde{H}-\tilde{\omega} I\right)}\}=0,
\end{align}
which is Eq.~\eqref{Eq: Imdet} with the following conditions:
\begin{subequations}
    \begin{equation}\label{Eq: ph1}
        \gamma_1\gamma_2\gamma_3+\gamma_1\kappa_{23}^2+\gamma_2\kappa_{13}^2+\gamma_3\kappa_{12}^2=0,
    \end{equation}
    \begin{equation}\label{Eq: ph2}
        \gamma_1+\gamma_2+\gamma_3=0,
    \end{equation}
\end{subequations}
where the condition~\eqref{Eq: ph2} stipulates a balance between the total gain and 
loss of the system. We consider the first resonator's gain $\gamma_1 = -g$ and the other 
resonator's losses $\gamma_2=\alpha \gamma_3$, leading to
\begin{align} \nonumber
	\gamma_2 &= \frac{\alpha g}{1 + \alpha}, \\ \nonumber
	\gamma_3 &= \frac{g}{1+\alpha}.
\end{align}

\paragraph*{Exceptional point.}
From Eq.~\eqref{Eq: Reldet}, we see that a third-order EP arises when the following
conditions are met:
\begin{subequations}
    \begin{equation}\label{Eq: ep1}
        \kappa_{12}^2+\kappa_{13}^2+\kappa_{23}^2+\gamma_1\gamma_2+\gamma_1\gamma_3+\gamma_2\gamma_3=0,
    \end{equation}
    \begin{equation}\label{Eq: ep2}
        2\kappa_{12}\kappa_{13}\kappa_{23}=0,
    \end{equation}
\end{subequations}
where the condition~\eqref{Eq: ep2} can be satisfied only if one of the coupling 
coefficients is zero. For $\kappa_{13}=0$, we can solve Eqs.~\eqref{Eq: ph1} 
and~\eqref{Eq: ep1} to obtain 
\begin{equation}\label{Eq: k13=0}
    \kappa_{12} = g\sqrt{\frac{1+\alpha}{2+\alpha}}, \, \kappa_{23} = \frac{g}{(1+\alpha)\sqrt{2+\alpha})}.
\end{equation}
The last possible case is $\kappa_{23}=0$. In this case, Eqs.~\eqref{Eq: ph1} 
and~\eqref{Eq: ep1} do not have a solution.

In Ref.~\cite{SK2022}, the constraints needed for the emergence of EPs in the 
presence of symmetries were described, where an nth-order EP can emerge when 
$2(n - 1)$ real constraints are satisfied. Our goal is to study the emergence 
of higher-order EPs within the context of pseudo-Hermitian systems. 
To generate a stable nth-order EP, both the pseudo-Hermiticity and  
EP conditions need to be satisfied, where $n$ eigenvalues and their corresponding 
eigenstates coalesce. For example, Eqs.~\eqref{Eq: ph1} and \eqref{Eq: ph2} define
the pseudo-Hermiticity condition, while Eqs.~\eqref{Eq: ep1} and \eqref{Eq: ep2}
ensure the existence of a third-order EP. Altogether, four constraints need to be 
satisfied, which is consistent with the previous result~\cite{SK2022}. 

In general, the feasibility of realizing higher-order EPs in one-dimensional space 
depends on the realization of the pseudo-Hermitian matrix. It is necessary that the 
conditions for both pseudo-Hermiticity and EP be satisfied. For instance, the 
condition in Eq.~\eqref{Eq: ep2} can only be satisfied if one of the coupling 
coefficients is zero. Yet, for $\kappa_{23} = 0$, the system becomes infeasible, as 
Eqs.~\eqref{Eq: ph1} and \eqref{Eq: ep1} no longer have a solution. This illustrates
the delicate balance needed to generate a higher-order EP in practical systems.

In the following, we study two distinct physical perturbation scenarios, each with 
its potential applications in threshold sensing, and present a circuit system with a 
non-injective structure that is not suitable for threshold sensing. 

\subsection{Mutual coupling $\kappa_{12}$ as a perturbation} \label{subsectionA}

\begin{figure} [ht!]
\centering
\includegraphics[width=\linewidth]{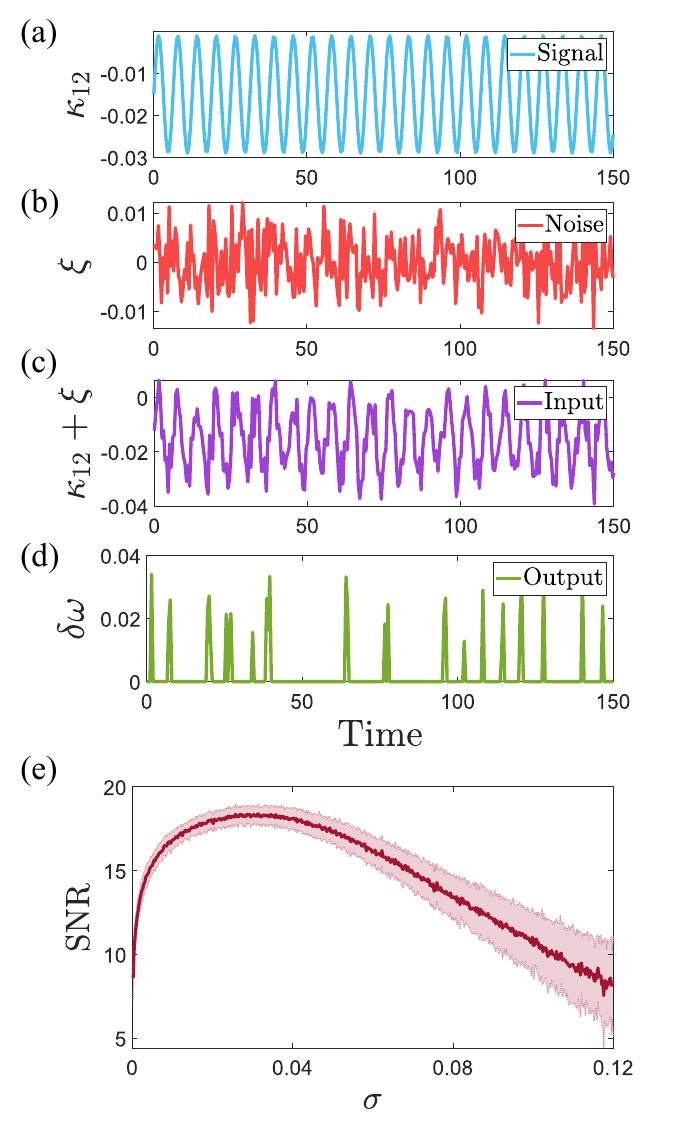}
\caption{Emergence of a stochastic resonance through a stochastic EP as a sensory 
threshold. (a) Input sinusoidal signal $\kappa_{12}(t)$ of amplitude 
$\kappa_b + \kappa_s = 0.099$. (b) Gaussian white noise of amplitude $\sigma=0.03$. 
(c) The noisy input signal. (d) Output of the system, a sequence of the pulses that 
appear randomly in time. (e) SNR as a function of the noise amplitude $\sigma$.
This smooth and broad SNR curve is obtained by averaging 1000 independent realizations 
of the process and the shaded area indicates the standard deviation. Other parameter 
values are $\alpha=50$, $\gamma = 0.1$, and $\kappa_{13} = 0$.}
\label{fig:threshold}
\end{figure}

For $\kappa_{13}=0$, Eq.~\eqref{Eq: Reldet} leads to three eigenfrequencies that 
evolve smoothly as a function of the perturbation $\kappa_{12}$. 
Figures~\ref{fig:coupled_system}(b) and \ref{fig:coupled_system}(c) show the real and 
imaginary parts of the eigenfrequency, respectively. It can be 
seen that, when $\kappa_{12}$ is perturbed, the response of the three coupled RLC 
resonators has a branch structure. Before the EP, ${\rm Re}\{\delta \omega \}=0$ and the 
system is insensitive to the input. At the bifurcation point $\kappa_{EP}$, all 
eigenfrequencies and their corresponding eigenvectors coalesce, and a small added 
perturbation can abruptly induce a frequency shift and result in a strongly nonlinear 
response. Now consider a time-varying inductive coupling parameter $\kappa_{12}(t)$, where
the input signal can be decomposed as $\kappa_{12}(t) = \kappa_b + \kappa_s sin(\nu t)$ 
with $\kappa_b$ the dc part of the coupling strength, $\kappa_s$ the amplitude of the 
oscillatory part, and $\nu$ the angular frequency. To investigate stochastic resonance
in such a structure, we consider white noise with standard deviation $\sigma$ added 
to the time-varying coupling $\kappa_{12}(t)$, where $\kappa_b+\kappa_s\leq\kappa_{EP}$. 
For $\kappa_{12}(t) \leq \kappa_{EP}$, a stochastic EP arises as a sensory threshold, 
resulting in an intermittent output of random frequency shift, as shown in 
Fig.~\ref{fig:threshold}(d), with a spectral peak at the signal angular frequency $\nu$. 

The effect of the stochastic EP can be quantified by the SNR of the output, which is the 
ratio of the signal power to the background noise power. To calculate the SNR 
statistically, we need the power spectral density - the Fourier transform of the 
autocorrelation function of the output. The power of the signal is proportional to the 
peak height of the power spectrum at the frequency of the time-varying input. Since all 
the spectrum other than the peaks is the background noise, the power of the noise is the 
sum of the rest of the spectral densities at the other frequencies. 
Figure~\ref{fig:threshold}(e) shows that the SNR of the system has a broad 
peak in $\sigma_c > 0$. Initially, as $\sigma$ increases from zero, the SNR rapidly 
increases, suggesting that the noise paradoxically enhances the performance of the 
sensor. This observation holds up to an optimal level of the noise $\sigma_c$, where 
the SNR gradually decreases beyond this point. This result indicates that stochastic 
high-order EP not only amplifies the sensitivity to perturbations but also contributes 
to an overall improvement in the system performance, which is characteristic of a 
stochastic resonance. The remarkable feature is that the SNR initially increases rapidly 
with the noise amplitude and, after reaching the maximum at an optimal noise level, the 
SNR decreases slowly. This gives rise to a wider range of the noise amplitude around 
the optimal noise to achieve a relatively large SNR. The phenomenon can be exploited 
for practical applications of sensors such as wearable sensors in which a stochastic 
EP arising from physiological motion overcomes the negative effect of noise, resulting 
in more accurate tracking of a person’s vital signs~\cite{li2023stochastic}. 

\subsection{Capacitive Perturbation $\varepsilon$}

\begin{figure} [ht!]
\centering
\includegraphics[width=\linewidth]{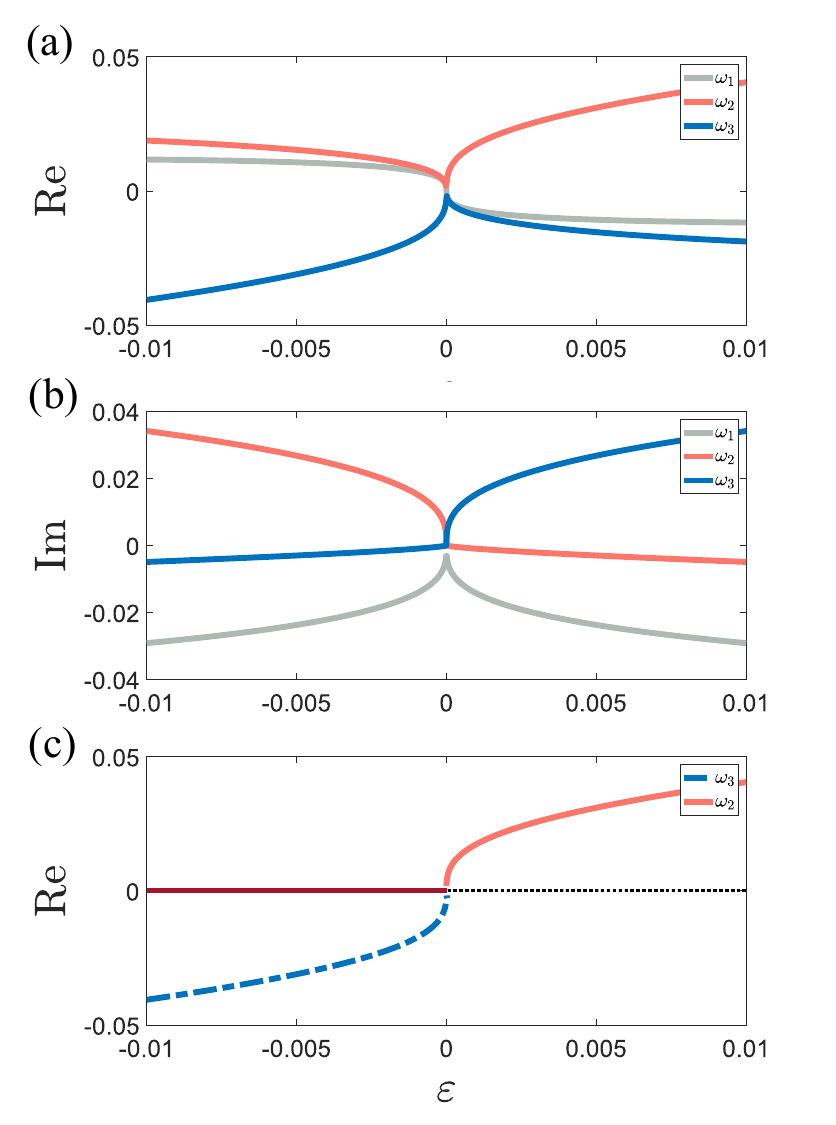}
\caption{Effect of a capacitive perturbation on the EP dynamics. Shown are the 
(a) real and (b) imaginary parts of the eigenfrequency evolution of the 
system~\eqref{Eq: schrodinger} as a function of the perturbation $\varepsilon$ for 
$\alpha=1$, $\gamma = 0.1$, $\kappa_{13} = 0$, and 
$\kappa_{23} = (1+\alpha)^{3/2}\kappa_{12}$.
(c) Reflection spectrum of the system in response to a capacitive perturbation.
Shown is the eigenfrequency shift in response to the perturbation $\varepsilon$.}
\label{fig:capacitive}
\end{figure}

Consider a capacitive perturbation $\varepsilon$ applied to the relay resonator at EP,
where $\kappa_{13}=0$, $\kappa_{12}$ and $\kappa_{23}$ take on values as in 
Eq.~\eqref{Eq: k13=0}. The characteristic equation of the perturbed system is 
\begin{align} \nonumber
\det{\left(\hat{H}-\varepsilon H_{e}-\hat{\omega} \mathbb{I}\right)}=0, 
\end{align}
where $H_e$ is a $3\times 3$ matrix that has one nonzero element on the second entry 
of the main diagonal. We have
\begin{equation}\label{Eq: chareq2}
    \hat{\omega}^3 - \varepsilon \hat{\omega}^2 + i \varepsilon \frac{\alpha g }{1+\alpha} \hat{\omega} - \varepsilon \frac{g^2}{1+\alpha} = 0.
\end{equation}
Fixing $\alpha=1$ and $g=0.1$, the real and the imaginary parts of the 
eigenfrequency can be obtained by solving Eq.~\eqref{Eq: chareq2}, as shown 
in Fig.~\ref{fig:capacitive}(a) and \ref{fig:capacitive}(b), respectively. It can be seen 
that any small perturbation applied to the EP (the bifurcation point of the system) 
gives rise to three different complex numbers, among which only one has physical 
significance: the eigenfrequency associated with a higher spectral resolution, in 
close relation to having a higher $Q$ factor or a narrower linewidth splitting 
${\rm Im}\{\delta \omega \}$. This suggests the eigenfrequency with the smallest magnitude 
of the imaginary part as the feasible choice for sensing applications, corresponding 
to the green curve $\hat{\omega}_{1}$ for $\varepsilon < 0$ and the red curve 
$\hat{\omega}_{2}$ for $\varepsilon > 0$ in Fig.~\ref{fig:capacitive}. The eigenfrequency 
shift in response to the perturbation $\varepsilon$ at EP exhibits a monotonic pattern, 
where the response of the system corresponding to one side of EP can selectively be 
filtered out, e.g., the green dot-dashed part of the response in 
Fig.~\ref{fig:capacitive}(c). In this case, the system has a saddle point 
$\varepsilon_{EP}$. The system's response to perturbations smaller than EP is 
unresponsive to the input (it is filtered out) and a small perturbation larger than EP 
can abruptly induce a frequency shift leading to a highly nonlinear response.

\begin{figure} [hb!]
\centering
\includegraphics[width=\linewidth]{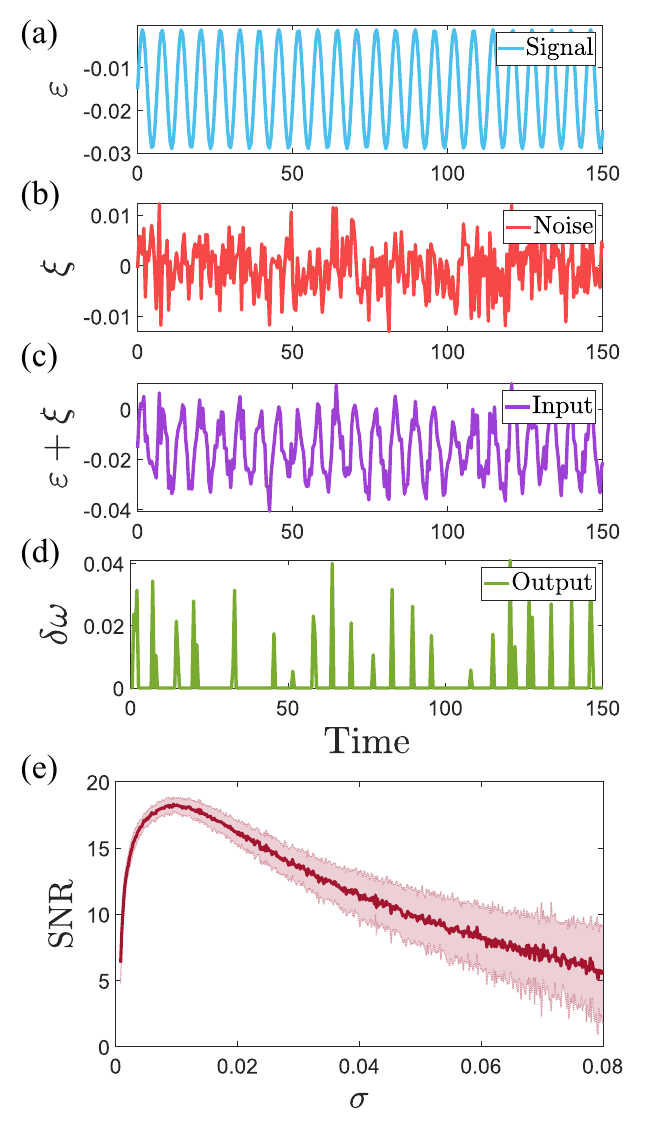}
\caption{Emergence of a stochastic resonance with a capacitive perturbation, where
a stochastic EP functions as a sensory threshold. The system is described by 
Eq.~\eqref{Eq: schrodinger}. (a) A time-varying sinusoidal input signal $\varepsilon(t)$
of the amplitude $\varepsilon = -0.01$. (b) Gaussian white noise of amplitude 
$\sigma=0.03$. (c) The noise input signal. (d) The output signal - a sequence of 
random pulses. (e) SNR versus the noise amplitude, which is indicative of a 
stochastic resonance. Similar to the resonance behavior in Fig.~\ref{fig:threshold}(e),
the SNR initially rises quickly with the noise amplitude and then decreases slowly
after reaching a maximum, giving rise to a relatively wide interval of the noise
amplitude for achieving a large SNR. Other parameter values are $\alpha=1$, 
$\gamma = 0.1$, and $\kappa_{13} = 0$.}
\label{fig:SR_capacitive}
\end{figure}

We now study the case where the input signal is the time-varying coupling parameter: 
$\varepsilon(t) = \varepsilon_b + \varepsilon_s \sin(\nu t)$, where $\varepsilon_b$ is the 
dc part of the coupling strength, $\varepsilon_s$ the amplitude of the oscillatory part, 
and $\nu$ is the angular frequency, and $\varepsilon_b+\varepsilon_s\leq\varepsilon_{EP}$,
as shown in Fig.~\ref{fig:SR_capacitive}(a). To induce a stochastic resonance, we add 
Gaussian white noise of amplitude $\sigma$, as shown in Fig.~\ref{fig:SR_capacitive}(b),
to $\varepsilon(t)$. The resulting noisy input signal is shown in 
Fig.~\ref{fig:SR_capacitive}(c), where $\varepsilon(t) \leq \varepsilon_{EP}$. The output 
of the system is intermittent, as shown in Fig.~\ref{fig:SR_capacitive}(d), corresponding 
to a random frequency shift with a spectral peak at the signal angular frequency $\nu$.
Figure~\ref{fig:SR_capacitive}(e) shows the the SNR of the system versus the noise 
amplitude. The SNR first rises quickly with the noise, reaches a maximum at the optimal 
noise amplitude $\sigma_c$, and then decreases slowly afterwards. This is indicative of a 
stochastic resonance in the presence of a stochastic high-order EP, signifying an 
enhancement in the system's performance assisted by noise in a wide range.

\subsection{A PT-symmetric electrical circuit}

\begin{figure} [ht!]
\centering
\includegraphics[width=\linewidth]{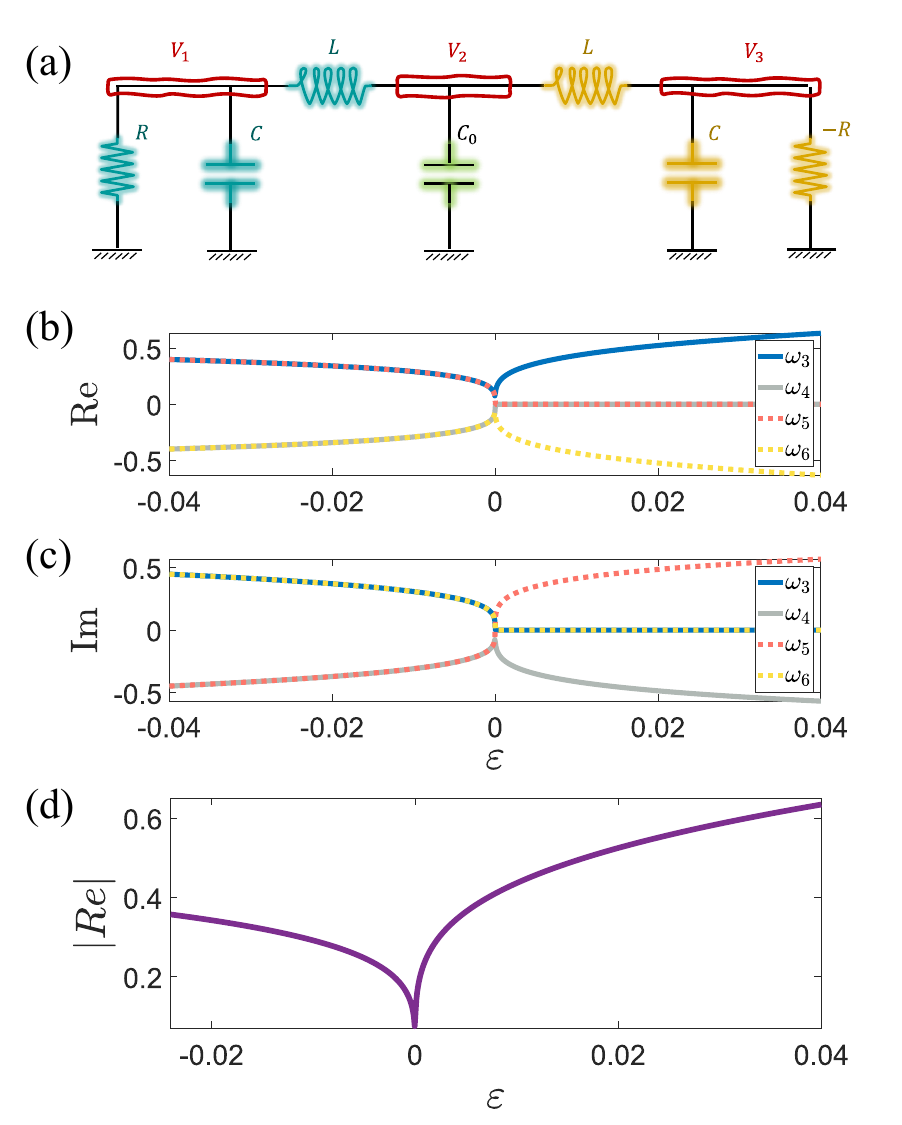}
\caption{A PT-symmetric sensing resonator with the negative impedance that can be 
realized with an operational amplifier. The system is described by 
Eq.~\eqref{Eq: 6-ordered}. (a) The circuit diagram. (b,c) Real and imaginary parts 
of eigenfrequency as a function of perturbation $\varepsilon$ to the parameter $\mu$.  
(d) Eigenfrequency shift in response to the perturbation $\varepsilon$. Other 
parameters are $\mu_0=(-1+\sqrt{5})/4$ and $\gamma = (1+\sqrt{5})/2$.}
\label{fig:circuit}
\end{figure}

We consider a PT-symmetric electrical circuit based on a sixth-order EP, which 
is composed of two LC resonators parallel with a resistor coupled with a grounded 
capacitor $C_0$, as shown in Fig~\ref{fig:circuit}(a). Using the current Kirchhoff’s 
law with normalization $\tau = \omega_0t$, we have that the resonant frequency 
is $\omega_0= 1/\sqrt{LC}$, the intrinsic loss or gain rate of the $LC$ resonator 
is $\gamma=R\sqrt{{C}/{L}}$, and the coupling coefficient between the two resonators 
is $\mu=C/C_0$. The voltages at various nodes of the circuit can be recast into 
the Schr\"{o}dinger-type equation as:
\begin{subequations}\label{Eq: 6-ordered}
    \begin{equation}
        i\dot{\Phi} = H \Phi,
    \end{equation}
    \begin{equation}
        H = i\begin{pmatrix}
            0 & 0 & 0 & 1 & 0 & 0\\
            0 & 0 & 0 & 0 & 1 & 0\\
            0 & 0 & 0 & 0 & 0 & 1\\
            -1 & 1 & 0 & -\gamma & 0 & 0\\
            \mu & -2\mu & \mu & 0 & 0 & 0\\
            0 & 1 & -1 & 0 & 0 & \gamma
        \end{pmatrix}.
    \end{equation}
\end{subequations}
The eigenfrequencies can be found through the associated characteristic equation 
${\rm det}(H-\omega \mathbb{I})=0$:
\begin{equation}\label{Eq: eq3}
    \omega^2[\omega^4+\left(\gamma^2-2-2\mu\right)\omega^2+1+2\mu-2\mu\gamma^2]=0,
\end{equation}
where $\mathbb{I}$ is the $6\times 6$ unity matrix. Solving Eq.~\eqref{Eq: eq3} 
leads to six eigenfrequencies as:
\begin{align}
    \begin{split}
        &\omega^*_{1,2} = 0,\\
        &\omega^*_{3-6} = \pm \frac{1}{\sqrt{2}} \sqrt{2+2\mu-\gamma^2 \pm \sqrt{\gamma^4 + (4\mu-4) \gamma^2 +4\mu^2}}.
    \end{split}
\end{align}
Figures~\ref{fig:circuit}(b) and \ref{fig:circuit}(c) show $\omega^*_{3-6}$ versus the 
perturbation $\varepsilon$ to the parameter $\mu$. It can be seen that the high-order EP 
at $\omega^*=0$ holds for $\gamma_{EP} = (1+\sqrt{5})/2$ and $\mu_{EP}=(-1+\sqrt{5})/4$.
For perturbation $\varepsilon < \varepsilon_{EP}$, all the eigenfrequencies share the same 
absolute imaginary parts (${\rm Im}\{ \omega \}$) with different real parts, resulting in 
the eigenfrequency shifts both to the right and to the left associated with the  
reflection coefficient of the system. For $\varepsilon > \varepsilon_{EP}$, the 
eigenfrequencies corresponding to the smallest ${\rm Im}\{ \omega \}$ are highlighted 
in yellow and green, which share the same absolute real parts. This implies that 
the reflection coefficient of the system has eigenfrequency shift both to the right 
and to the left. Consequently, the response of the system to perturbation 
corresponding to the smallest ${\rm Im}\{ \omega \}$ belongs to that of a non-injective 
structure with a bifurcation point at $\varepsilon_{EP}$. Figure~\ref{fig:circuit}(d) 
shows the absolute values of the frequency shift with respect to the applied 
perturbation. Due to the symmetric nature of the system's response, the sensor 
illustrated in Fig.~\ref{fig:circuit} is not a suitable choice for sensory threshold 
applications. Instead, a potential application lies in enhancing the system's 
sensory response rather than serving as a sensory threshold. A key factor is that 
the system's response to perturbations, whether they are of higher or smaller values 
than the EP, cannot be reliably distinguished~\cite{xiao2019enhanced}.

In this example of a PT-symmetric electrical circuit, there is a symmetric
response to perturbations. Regardless of whether the perturbation is above or below
the EP, the eigenfrequency shifts to the right or the left, depending on the reflection
coefficient of the system. Due to this symmetry, the sensor presented in Fig.~9
cannot serve as a threshold sensor because it lacks the ability to distinguish
between higher and lower perturbations. Instead, such systems are more suited for
applications where enhancing overall sensitivity is the goal, rather than serving
as a threshold sensor.

\section{Discussion} \label{sec:discussion}

We have investigated the possibility of exploiting pseudo-Hermitian systems for sensing 
applications, focusing on the dynamics about a high-order EP. The frequency response 
of the sensory system with a high-order EP to perturbations can be categorized into three 
scenarios: branch, monotonic, and non-injective. In each case, the splitting in the 
real part of the frequency reveals EPs as a critical point around which the system 
exhibits enhanced sensitivity to perturbations. A high-order EP can enhance 
sensitivity, rendering it desired for sensory threshold applications, where the system 
can leverage the presence of noise to improve its performance. We have demonstrated that 
the interplay among the exceptional point, perturbation as input signal, and noise
leads to a stochastic resonance. This stochastic resonance associated with a high-order 
EP has one appealing feature. As indicated in Figs.~\ref{fig:threshold}(e) and 
\ref{fig:SR_capacitive}(e) for two distinct types of perturbations (input signals), the 
SNR versus the noise amplitude exhibits a broad maximum. This means that a precise tuning 
of the noise amplitude is not required, as there exists a range of the amplitude in 
which the SNR maintains at a high value. The phenomenon of high-order EP induced 
stochastic resonance not only underscores the practical applicability but also extends 
the boundaries of potential applications in signal processing and sensor technology. 

The energy or frequency splitting at an EP
follows an exponential relationship with the strength of the perturbation,
expressed as $\epsilon^{1/(n-1)}$, where $n$ is the order of the EP. This relation 
implies a significantly higher sensitivity compared to that associated
with a diabolic point found in conventional sensors. In our work, we extended this 
sensitivity to scenarios where the input signal is weak and typically set at or 
below the sensory threshold. Under such conditions, without noise, no detectable 
output would be expected. However, due to the system’s extreme sensitivity at the EP,
even the intrinsic noise of the system can act as a perturbation, enabling the
detection of an output. The SNR values depicted in Figs.~\ref{fig:threshold}
and \ref{fig:SR_capacitive} reflect this 
phenomenon, demonstrating that there exists an optimal noise strength where the SNR 
is maximized. This outcome is directly related to the exponential sensitivity of the 
EP to perturbations, including noise. At low noise levels, the input signal remains
undetectable, while at higher noise levels, the system becomes overwhelmed by
noise. As a result, there is an optimal noise strength that enhances the SNR by
leveraging the EP's sensitivity to weak perturbations. For a fair comparison,
the original SNR value without any noise or perturbations can be considered as 
a baseline value, where the system would exhibit no detectable signal in the output 
due to the input signal being below the threshold. The results in 
Figs.~\ref{fig:threshold} and \ref{fig:SR_capacitive}
thus demonstrate how the introduction of noise allows the system to surpass this 
baseline and exhibit enhanced detection capability.

It is worth noting that the extended (monotonic) structure is particularly appealing
for real-world sensing applications. It can arise in systems with PT or pseudo-Hermitian 
symmetry, where broader experimental platforms are available compared with the 
branch-type of structures. For example, in an optomechanical 
accelerometer~\cite{kononchuk2020orientation} consisting of a pair of Fabry-P\'{e}ot 
multilayer cavities with loss and gain cavity, respectively, a PT symmetric system
with a second-order EP can be constructed on a silicon platform. In this system, the 
positive (negative) acceleration results in the right (left) hand shift direction in 
spectra, which is exactly what the monotonic structure exhibits. For pseudo-Hermitian 
systems with a third-order EP~\cite{xiong2021higher} in cavity optomechanics, the 
eigenvalue spectrum versus the detuning disturbance displays the extended structure 
since only the zero imaginary part can be distinguished and observed in the experiment. 
As a result, with the monotonic structure, capacity as disturbance can then be applied 
to an accelerometer for a wide range of 
applications~\cite{kononchuk2020orientation,krishnan2007micromachined,shindo2002large,bao2000micro}, including navigation devices, gravity gradiometry, earthquake 
monitoring~\cite{wu2017nano,li2016novel}, airbag deployment sensors in automobiles, 
and consumer electronics protection. In addition, the capacity can also be designed as 
a hypersensitive microfluid speed sensor (temperature sensor)~\cite{malmberg1956dielectric,xiao2019enhanced} and pressure sensor~\cite{chen2014high,chen2018generalized}.

\section*{Acknowledgments}
This work was supported by the Air Force Office of Scientific Research under Grant 
No.~FA9550-21-1-0438 and by the Office of Naval Research under Grant No.~N00014-24-1-2548.

%\bibliographystyle{agsm}
%\bibliography{HOEP}

%apsrev4-2.bst 2019-01-14 (MD) hand-edited version of apsrev4-1.bst
%Control: key (0)
%Control: author (8) initials jnrlst
%Control: editor formatted (1) identically to author
%Control: production of article title (0) allowed
%Control: page (0) single
%Control: year (1) truncated
%Control: production of eprint (0) enabled
%
\end{document}